\documentstyle[amssymb,preprint,aps]{revtex}

\tightenlines

\begin{document}
\title{A Cosmic Lattice as the Substratum of Quantum Fields}
\author{Helio Fagundes}
\address{Instituto de F\'{\i}sica Te\'{o}rica\\
\ \ Universidade Estadual Paulista \\
S\~{a}o Paulo, SP 01405-900, Brazil\\
E-mail: helio@ift.unesp.br}
\date{Publ. in Brazil. J. Phys. {\bf 19}, 27 [1989]. Sequel in preparation.}
\maketitle

\begin{abstract}
A cosmology inspired structure for phase space is introduced, which leads to
finitization and lattice-like discretization of position and momentum
eigenvalues in a preferred, cosmic frame. Lorentz invariance is broken at
very high energies, at present inaccessible. The divergent perturbation
terms in quantum electrodynamics become finite and small; this could become
a requirement leading to model restrictions in other perturbative theories.
So the very success of the usual renormalization procedures is simply
explained by their finitization, and is viewed as indicating the reality of
the lattice.
\end{abstract}

\section{The boundless but finite cosmic lattice}

The idea of assigning a fundamental length to physical space has a long
history. A large part of the literature can be traced from the work of
Gudder and Naroditsky \cite{one}. This notion arises from dissatisfaction
with the way renormalizadion operations are performed in quantum field
theory - both the nonrigorous and the too-rigorous methods. A lattice
structure for space or spacetime is an obvious \ remedy for the worst
problem, ultraviolet divergences, but people have been discouraged by the
prospect of Lorentz and rotational non-invariance. Recently some work has
appeared, which faces just this possibility. Thus Nielsen and collaborators
have developed both a quantum electrodynamics (QED) and a Yang-Mills theory
that violate Lorentz invariance \cite{two}, and Zee \cite{three} has gone so
far as suggesting experimental verification of such a breakdown. Some time
ago Wheeler \cite{four} pointed out the possibility of a breakdown of the
spacetime concept itself on the scale of Planck's length.

In this paper I introduce upper limits for both measurable length and
measurable momentum, as a means of reaching a kind of lattice structure for
phase space. The conclusion will be that the perturbative series of QED
works so well because its terms become, in a quasi-invariant sense to be
defined, convergent to small numbers. The idea has a cosmological
inspiration: a Friedmann-Robertson-Walker (FRW) model with flat space
sections \cite{five} has the local metric $ds^{2}=c^{2}dt^{2}-b^{2}(t)d{\bf x%
}^{2}$. Its global spatial topology may be assumed to be that of a flat
torus $T^{3}$, which is an identification space isometric to the quotient
space $E^{3}/\Gamma $, where $E^{3}$ is Euclidean space and $\Gamma $ is a
group generated by finite translations in the three directions. For the
development and motivations of this idea in cosmology see the list of refs. 
\cite{six}, in particular Ellis and Schreiber's recent work.

The expansion factor in this metric is $b(t)=(t/t_{0})^{2/3}$, where $%
t_{0}\sim 10^{10}$ yr is the age of the of the universe. Since $\left(
db/dt\right) (t_{0})\sim 10^{-18}\ \sec ^{-1}$, let us put $b(t)=1$ in $%
ds^{2}$ for the discussion of laboratory physics (adiabatic approximation).
We are left with a locally Minkowskian spacetime, whose $T^{3}$ spatial
sections may be obtained from a cube of side $L$, by the identification
(``gluing'') of opposite faces. Therefore $\sqrt{3}L$ will be a maximum
distance in space, and I take $L=c/H$, where $H\approx 75$ km sec$^{-1}$Mpc$%
^{-1}$ is Hubble's constant.

Now I make the crucial assumption that momentum space (relative to the above
cosmic frame) has the same $T^{3}$ topology as configuration space, so that
the corresponding phase space is the product manifold $T^{3}\times T^{3}$.
Besides providing an upper limit for momentum, this sort of duality appears
to the author as a more reasonable way of assuring discretization of
position than just drawing from crystal structure analogy, with its
``neo-ether'' connotation. The flat torus for momentum is obtained by
identifying opposite faces of a cube of side $P=2\pi \hslash /a$, where $%
a=(G\hslash /c^{3})^{1/2}=1.61\times 10^{-33}$ cm is Planck's length.
Incidentally, a cutoff for momentum is quite reasonable, for otherwise the
energy of a single virtual particle can exceed the total mass of the
observable universe. It is also in agreement with Wheeler's idea cited
above: if spacetime breaks down at Planck's length, so must the validity of
momenta greater than $P$. (Of course the $T^{3}\times T^{3}$ topology
postulate may come to be seen as a phenomenological assumption, if and when
future developments lead to a more general theoretical framework, where
phase space would be perceived as an approximation suitable for 1988 physics
- like the idea of orbit that, from a quantum mechanical viewpoint, is seen
as an approximation suitable for the classical limit.)

In quantum mechanics we are used to box quantization with periodic boundary
conditions imposed by convenience. But in the above defined cosmic tori
these conditions are just the natural ones. So we get eigenvalues $%
x^{k}=n^{k}a$ and $p^{k}=n^{k}(\pi \hslash /Na)$, $k=1,2,3$, $-N\leq
(n^{k}=\ ${\normalsize integer}$)\leq N=L/2a\approx 3.8\times 10^{60}$. Thus
space becomes a very large box, which is both finite and boundless, with
discrete eigenvalues for particle position and momentum. This suggests that
we call it a {\it cosmic lattice} (CL), but note it is an abstract, not a
granulated, crystal-like lattice.

\section{Lorentz quasi-invariance}

What about Lorentz invariance? First, the preferred status of the CL frame
should not cause much surprise. It is the home frame of our cosmos, similar
to the comoving system of Einstein-de Sitter's cosmology, the $2.7$ K
radiation providing its concrete realization (except for spatial
orientation; see Gott \cite{six} for an explanation on how the apparent {\it %
isotropy of cosmic observations} is preserved despite the loss of global
invariance for rotations). Second, let us define the composition of
4-momenta $p_{1}^{\mu }$ and $p_{2}^{\mu }$ in the CL system. Setting $%
\hslash =c=1$, if $|p_{1}^{k}+p_{2}^{k}|\leq \pi /a$, then energy-momenta
are composed as usually: $p^{\mu }=p_{1}^{\mu }+p_{2}^{\mu }$. If $%
|p_{1}^{k}+p_{2}^{k}|>\pi /a$, then we add or subtract $2\pi /a$, so as to
obtain $p^{k}$ in the allowed range. This $p^{k}\ (\equiv p_{1}^{k}+p_{2}^{k}%
\mathop{\rm mod}%
2\pi /a)$ is defined to be the resultant. With $s\equiv
(p_{1}^{0}+p_{2}^{0})^{2}-({\bf p}_{1}+{\bf p}_{2})^{2}$, the resultant
energy is $p^{0}=(s+{\bf p}^{2})^{1/2}\leq p_{1}^{0}+p_{2}^{0}$. Therefore
energy-momentum conservation is only conserved in collisions if ${\bf p}_{1}+%
{\bf p}_{2}$ is a CL momentum eigenvalue. But this is hardly a constraint,
since laboratory momenta are far from our limit, $\sqrt{3}\pi /a\sim 10^{20}$
GeV/c.

Thirdly, Lorentz transformations, including rotations, are performed as
usually, but the limits of $p^{k}$ in an arbitrary frame are derived from
those in the CL frame. Thus if a system has velocity ${\bf \beta }=
(v,0,0),v>0$, and no rotation with respect to the CL, then 
\[
p_{\max }^{1}=\gamma \left[ \frac{\pi }{a}-v\left( m^{2}+\frac{\pi ^{2}}{%
a^{2}}\right) ^{1/2}\right] \approx \frac{\pi }{2a\gamma }\ 
\]
and 
\[
p_{\min }^{1}=\gamma \left[ -\frac{\pi }{a}-v\left( m^{2}+\frac{\pi ^{2}}{%
a^{2}}\right) ^{1/2}\right] \approx -\frac{2\gamma \pi }{a}\ , 
\]
for large $\gamma $. The limits for $p^{2}$, $p^{3}$ remain unaltered. If we
now rotate this frame, the physics will be invariant if all relevant momenta
are smaller than $\pi /2a\gamma $. Again, in laboratory situations we do not
have to worry about these limits. Summarizing, Lorentz and rotational
invariance are preserved if we restrict ourselves to laboratory energies and
to\ a theoretical range of frames - say, $\gamma <10^{10}$ with respect to
the CL, which guarantees invariance up to $\sim 10^{9}$ GeV in the moving
frame. I shall refer to this restricted meaning as {\it Lorentz
quasi-invariance.}

\section{Finite renormalization}

The practice of renormalization on QED has been so strongly associated with
the removal of infinities that the fundamental meaning of the former became
blurred. See, for example, Schweber's \cite{seven} warning against this
tendency. Actually the aim of renormalization is to combine some
unobservable parameters of a basic model into observable ones, so that the
renormalized model is expressed in terms of the latter. The advantage of
this process is obvious when one considers that the purpose of theoretical
models is to represent experimental facts. Consider mass renormalization in
QED: we write $m=m_{0}+\delta m$, and say that $m$ is the experimental mass
of the electron. But the underlying formalism suggests that we also
interpret $m_{0}$ and $\delta m$ anyway, as bare mass and the effect of
virtual photons always surrounding the electron respectively. It seems to
the author that if $m_{0}$ and $\delta m$ can be made finite, so much the
better: their interpretation is reinforced, and we might even think of
making them observable - as when one tries to assign mass differences in
isospin multiplets to electromagnetic interactions \cite{eight}. (See,
however, \cite{nine}.)

Therefore my program is not to abandon renormalization, but rather to make
it step-by-step finite \cite{ten}. I will essentially follow the established
formalism with a few adaptations: (a) integrals are in principle replaced by
sums over the CL, but in practice the latter are approximated by integrals
that formally resemble the original ones but are now finite and small (and
justifiably so); (b) the calculations are preferably performed in the CL
reference frame, which is the natural system in this context, just like a
Sun-centered system is natural for planetary astronomy (this naturalness can
of course be formalized); (c) restricted Lorentz transformations, as
discussed in Sec. II, are seen to hold for the results of calculations.

Let us examine some problems of perturbative field theory in terms of the
above ideas. The great success of the usual formalism suggests that we try
to keep its analytical basis, rather than for example switching to
difference equations \cite{one}, \cite{eleven}. This is physically
reasonable, since the scale of Planck's length is so much finer than that of
currently observable processes. Therefore I will here assume minimum
departures from established analytical expressions. For comparison with
standard results I will rely on Itzykson and Zuber's textbook \cite{twelve},
henceforth referred to as (IZn), where n is the page number.

The infrared catastrophes will be transformed into finite contributions
(since the minimum energy of a massless particle is $\pi /Na$, not zero),
and if these are still too large they may be dealt with as usually, e. g. as
in (IZ334).

Consider now charge renormalization in QED. The notation below is adapted
from (IZ319ff). The one-loop contribution to vacuum polarization, after use
of Feynman's trick $(ab)^{-1}=\int_{0}^{1}dz[az+b(1-z)]^{-2}$, is \cite%
{thirteen} 
\begin{equation}
\omega _{\rho \nu }(k)=-4e^{2}\int_{0}^{1}dz(z-z^{2})\newline
\sum_{{\bf q\in }\text{CL}}\int_{-\infty }^{\infty }\frac{dq_{0}}{2\pi }%
\frac{(g_{\rho \nu }k^{2}-2k_{\rho }k_{\nu })(z-z^{2})-g_{\rho \nu
}(q^{2}/2-m^{2})}{\left[ q^{2}+k^{2}(z-z^{2})-m^{2}+i\epsilon \right] ^{2}}\
.
\end{equation}%
This expression is well defined, so it can be safely simplified by gauge
invariance, which leads to 
\begin{equation}
\omega _{\rho \nu }(k^{2})=-i(g_{\rho \nu }k^{2}-k_{\rho }k_{\nu })\omega
(k^{2})\ ,
\end{equation}%
with \ 
\[
\omega (k^{2})=2e^{2}\int_{0}^{1}dz(z-z^{2})\sum_{{\bf q\in }\text{CL}}|{\bf %
q}^{2}+m^{2}(z-z^{2})|^{-3/2}.
\]%
Eq. (2) is Lorentz quasi-invariant, in the above defined sense. Although
calculated in the CL frame, $\omega (k^{2})$ is an invariant - like, say,
the contribution of vibrational energy to the mass of a crystal.
Approximating \cite{fourteen} the sum over the cosmic box by an integral
over a ball of radius $\pi /a$, and neglecting positive powers of $ma$, I
obtained for $k^{2}<4m^{2}$, 
\[
\omega (k^{2})=\frac{\alpha }{3\pi }\left[ \ln \left( \frac{2\pi }{ma}%
\right) ^{2}-2+\frac{k^{2}}{5m^{2}}\ ...\ \right] .
\]%
Hence $Z_{3}=[1+\omega (0)]^{-1}=0.925$ and $e=0.962e_{0}$. The Uehling term
is the same as in (IZ327).

Similarly, I got for mass renormalization, to the same order, 
\[
\frac{m}{m_{0}}=1+\frac{3\alpha }{4\pi }\left[ \ln \left( \frac{2\pi }{ma}%
\right) ^{2}-\frac{1}{3}\right] =1.185\ ,
\]%
and, in Feynman's gauge, 
\[
Z_{2}^{-1}-1=\frac{\alpha }{4\pi }{\left[ \ln \left( \frac{2\pi }{ma}\right)
^{2}+2\ln \left( \frac{\pi }{Nma}\right) ^{2}+3\right] ,}
\]%
hence $Z_{2}=1.160$. If we compare the above results with their counterparts
in (IZ325,334,335), we see that the logarithmic terms in the former can be
obtained from those in the latter if we replace $\Lambda $ by $2\pi /a$ and $%
\mu $ by $\pi /Na$. The author hopes to derive similar results for higher
order terms in QED, and possibly for other theories of fundamental
processes. The important immediate consequence of the achieved finitization
is that the terms of the perturbative series, which are normally understood
in a context of {\it formal} procedures, to \textquotedblleft extract
sensible results from apparently ill-defined expressions\textquotedblright
(IZ318), become {\it legitimized} as ordinary finite terms. Theories
satisfying this condition could be called {\it perturbatively renormalizable}%
, and this property might be a further guide for model building. (So, for
example, the perturbative treatment of $\lambda \phi ^{4}$ models might be
deemed unacceptable, because of its quadratic divergence in mass
renormalization.) As bonuses, the calculations become less dificult - the
\textquotedblleft naive prescription\textquotedblright (IZ374) of cutting
off large momenta becomes the natural one - and the meaning of the
renormalized Lagrangian gets a numerical foundation - compare (IZ345,346).
Interpreted in this light, the fact that renormalization theory has been so
successful can be invoked as an argument for the physicallity of the CL (or
some related concept). It remains to be seen whether this notion will be
tested, for example in proton decay, as suggested by Zee \cite{three}.

I am grateful to C. Rebbi for a private correspondence, to P. Leal Ferreira
for calling my attention to \cite{three}, and to many colleagues for
discussions on the problems of renormalization, in particular a long
conversation with H. Fleming. This work has been supported by the Brazilian
agencies FINEP and CNPq.


\begin{references}
\bibitem{ten} One of the discoverers of QED renormalization has written that
the rules for eliminating the infinities are like ``sweeping the dirt under
the rug''(cf. [8], p. 37). These rules can perhaps be justified by
sophisticated mathematics. Here instead I am suggesting a mathematically
simple and physically sensible way to avoid the occurrence of infinities.
\bibitem{fourteen} A better approximation would be to integrate over a cube
of side $2\pi /a$, but for this preliminary work I adopted the simpler
integration over a ball.
\end{references}
\end{document}